# Magnetization pinning in conducting films demonstrated using broadband ferromagnetic resonance

M. Kostylev[1], A.A. Stashkevich[3], A.O. Adeyeye[2], C. Shakespeare[1], N. Kostylev[1], N. Ross[1], K. Kennewell[1], R. Magaraggia[1], Y. Roussigné[3], and R. L. Stamps[1]

[1]*School of Physics, University of Western Australia, Australia*

[2]*Information Storage Materials Laboratory Department of Electrical and Computer Engineering, National University of Singapore, Singapore*

[3]*LPMTM CNRS (UPR 9001), Université Paris 13, 93430 Villetaneuse, France*

## Abstract

The broadband microstrip ferromagnetic resonance technique has been applied for detection and characterization of a magnetic inhomogeneity in a film sample. In the case of a 100nm thick Permalloy film an additional magnetically depleted top sub-layer, practically unidentifiable by the conventional ferromagnetic resonance setup, has been detected and characterized. These results have been confirmed by Brillouin light scattering spectroscopy revealing the fact that the optical properties of the additional sub-layer do not differ much from those of the bulk of the film. Subsequent characterization of a large number of other presumably single-layer films with thicknesses in the range 30-100nm using the same ferromagnetic resonance technique also revealed the same effect.

## Introduction

Applications of metallic ferromagnetic materials to the microwave frequency devices motivate interest in high-frequency dynamics of continuous and patterned ferromagnetic films. In the case of relatively low angle precession, magnetic dynamics manifests itself through bulk magnetic excitations known as spin waves (SW). Breakthroughs in the technology of single crystal yttrium iron garnet (YIG) ferrite films with extremely low losses at microwave frequencies suitable for effective signal processing, typically in delay lines in the frequency range from 2 GHz to 20 GHz,[1,2] has caused further development of this concept in the 1960s – 1980s. Although soft ferrite materials have found important applications in microwave engineering,[3] their frequency performance is severely handicapped by small values of the saturation magnetization. There is consequently a growing research effort in the field of microwave properties of soft ferromagnetic metals, such as Permalloy or iron, for applications to wide-band microwave signal processing.

In particular, it has been shown that despite relatively high insertion losses Permalloy films can be used in delay lines,[4] as well as in parametric amplifiers.[5] They are also regarded as potential candidates for magnetologic devices,[6,7] emerging as a new interesting field of application of propagating SW, not directly related to microwave signal processing. The Mach-Zehnder geometry proposed in Ref.8 and whose efficiency in the case of a YIG structure has been proved in Ref.9, can be regarded as a typical example of the basic physical principle underlying logic operations based on destructive and constructive SW interferences. However, since metal films are at least one order of magnitude thinner than their ferrite analogues, they allow creation of much smaller microwave and logic components.[8] Moreover sputtering technology employed in the fabrication of metal films, makes it much easier to incorporate magnetic layers into hybrid integrated structures, than in the case of conventional insulating ferrites prepared by liquid phase epitaxy.[10-12]

Finally, one should not underestimate another degree of freedom characteristic of metal films and entirely non-existent in conventional insulating ferrites: the presence of free charge carriers. The recent discovery of the Doppler frequency shift of a SW interacting with a current in a Permalloy film[13,14] has drawn attention to the importance of the above mentioned physical property.

The performance of microwave and logic components, employing Permalloy films, drastically depends mostly on two parameters: high frequency damping and pinning conditions. Although the high frequency absorption in Permalloy films has been investigated since the 1970s[15] this characteristic is still extensively discussed in recent publications.[16,17]

The same applies to the pinning conditions (see e.g. Ref.18) which, as is well known, change significantly the SW spectra.[19] In this paper we report detection of an intermediate thin magnetic sub-layer causing, via the so called dynamic pinning, noticeable changes in the dynamic behavior of the whole structure.[20] To ensure reliability of the experimental results, we have adopted a combined technique, employing two completely independent approaches: ferromagnetic resonance excited by an external wide micro-strip antenna and Brillouin Light Scattering (BLS) by thermal magnons.

In the first case we made use, as proposed in Ref.21, of the characteristic asymmetry of the two interacting microwave fields: the field of the exciting structure and that of the excited mode. In the complementary BLS approach we appealed to the dispersion characteristics of thermal magnons in a wide range of wave-numbers which is covered by this powerful technique. More specifically, we made use of the non-reciprocity of the Damon-Eshbach[22] SW excitations, propagating in an asymmetric ferromagnetic structure, as first reported in Ref.23.

## Experiment

Microwave absorption is a common tool in the study of high frequency properties of metallic ferromagnetic films and multilayers. Because of the sensitivity of some magnetic excitations to film surfaces and interfaces, this is a particularly useful technique for nondestructive studies of magnetic interfaces.

In our recent experimental[24] and theoretical[21] papers we have shown that magnetic metallic bi-layer films, tailor-made to optimize the efficiency of dynamic pinning, manifest a strong asymmetry of microstrip broadband FMR response with respect to layer ordering. More specifically, in Ref.24 we measured FMR responses from a series of Cobalt-Permalloy bi-layer films grown on Silicon substrates, prepared ad hoc to test the efficiency and reliability of FMR characterization of such structures: for all films the Cobalt (Co) layer thickness was kept at 10 nm, while the Permalloy (Py) thickness varied from sample to sample in the range from 40nm to 100 nm.

The FMR "signature" defined as the number and as shape of the absorption lines depends drastically on the overlap between the exciting high frequency field, produced by a microstrip and the magnetization profile $\mathbf{m}(y)$ in the excited spin wave mode. Both have a pronounced asymmetry across the film. However, while the asymmetry of $\mathbf{m}(y)$ is fixed with respect to the layer ordering, the external field can be applied from both sides by placing the sample on the microstrip differently: with the sample side down or with the substrate down. In the latter geometry, to ensure the exposure of the bi-layer to the exciting field, the width of the microstrip was made greater then the substrate thickness. Theoretical simulation (see Ref.21) based on a well defined sample geometry and on typical magnetic parameters of both metals gave excellent results.

In this paper, applying the same approach, we address an inverse problem: detection and characterization of a supplementary layer in a seemingly homogeneous Permalloy single-layer. The test sample, 100 nm thick, has been grown on a 500 μm thick Si substrates by standard DC magnetron sputtering.

In our broadband FMR experiments we use an Agilent N5230A PNA-L microwave network analyzer to apply a microwave signal to the samples and to measure magnetic absorption. As a measure of the absorption we use the microwave scattering parameter *S21* [25]. We keep the microwave frequency constant and sweep the static magnetic field *H* applied in the film plane and along the transducer to produce resonance curves in the form of *S21*(*H*) dependences. This is repeated for a number of frequencies. We also measure *S21* for the transducer with no sample ($S21_0(H)$) to eliminate any field-dependent background signal from the results. The results presented below are Re($S21(H)/ S21_0(H)$).

The driving current is applied to a section of microstrip. The sample sits on top of this microstrip transducer, whose width is 1.5 mm. The electromagnetic field in a TEM mode supported by the microstrip line is described by the Laplace equation. It falls down across the substrate sufficiently slowly to provide for efficient SW excitation in the unfavorable geometry when it is the substrate that faces the microstrip.

Preliminary experimental results of the FMR measurements signaling the presence of a second sub-layer are given in Fig.1. One can clearly see that the position of the sample with respect to the transducer (film facing the microstrip / substrate facing the microstrip), as it has been outlined in Refs.21, 24, changes drastically the response: the profile transforms from a single peak to twin peak curve. The single peak is traditionally identified, according to its field position, as the fundamental "Kittel" mode. Since the time of numerous classical cavity FMR studies,[26] observing only a fundamental mode is traditionally considered as an evidence of high homogeneity of film properties across the film thickness, in particular, the absence of surface spin pinning.[27]

Important here is the fact that when we place the sample with the substrate facing the wide microstrip transducer we detect a second intense response (Figure 1, dashed line). The latter attests to some sort of pinning on one of the interfaces. The problem is, however, to define on which surface of the main Py film the additional “pinning” layer is formed. In any case, it is plausible to suggest that this layer is magnetically depleted, either due to oxidation on the free surface or via inter-diffusion at the film-substrate interface.

To identify the nature of this sub-layer and to characterize its parameters extensive numerical simulations have been performed. Representative theoretical fits of experimental data are given in Fig.2. The following situations are illustrated: $f$ = 8 GHZ, $f$ = 12 GHZ with the substrate facing the transducer and $f$ = 15 GHZ with the film facing the transducer. By fitting responses for both film placements with the theory[21] one obtains the thickness $L_s$ and saturation magnetization $4\pi M_s$ for the sub-layer. The best fit is obtained for $L_s$=10nm and $4\pi M_s$=4000G (See Fig. 2.) Moreover, numerical simulations suggest that the depleted sub-layer is localized on the *free surface.* Similar evidence of the presence of a magnetically depleted sub-layer was also found for the second presumably single-layer film from the same series.

Further evidence of the presence of this layer was obtained by measuring thermal magnon dispersion with Brillouin light scattering (BLS) technique at the University Paris-Nord. The results obtained are illustrated in Fig.3, where experimental points corresponding to BLS peaks are superimposed on theoretical dispersion curves. This measurement revealed a considerable difference in frequencies for the Stokes and anti-Stokes BLS peaks, characteristic of a multi-layer with an asymmetric magnetic structure across the film.[23] This difference is well fitted by a non-reciprocal spin wave dispersion law for a bi-layer with the material parameters extracted from the FMR data above. It should be stressed that here, once again, we have supposed that the depleted layer is situated on top of the main one. Another significant feature pointing to the presence of dynamic pinning is the appearance of a pronounced dipole-exchange gap between the DE mode and SWRs, well reproduced both in theory and experiment. However, this peculiarity, while describing the extent of effective pinning, provides no clue as to at which surface it takes place.

## Discussion

One of the major points addressed in this paper is how to identify the surface at which the effective pinning takes place. Interestingly, in the case investigated in Ref.24 (a thick Py film with a thin "pinning" Co sub-layer) the larger amplitudes for the higher-order modes were detected for the pinning layer facing the transducer. In other words, in the case when the microstrip field penetrated the sample through *the thin supplementary sub-layer*. At the same time, in our case just the opposite was observed: higher modes were generated when the exciting magnetic field penetrated *through the main layer* (see Fig.1). The theory in Ref.21 gives indications that this can be related to the reversal of the magnetization profile in the two magnetic structures studied. In Ref.24 the thin pinning sub-layer has a higher saturation magnetization $4\pi M_s$ then the bulk of the sample, while in our case the depleted pinning sub-layer is characterized by a lower value of $4\pi M_s$.

Additional evidence in favor of this explanation, based on our numerical simulations, is presented in Fig.4. According to Ref.21 the absorption amplitudes for microstrip broadband ferromagnetic resonance depend on two major factors: the shape of the thickness profile of the total microwave magnetic field inside the sample and the feedback from the excited magnetization to the microstrip. The former determines how efficiently magnetization precession is excited; the latter determines how efficiently the signal of precessing magnetization is picked up by the transducer.

The total microwave field consists of the microwave field of the microstrip transducer, the field of eddy currents induced in the sample, and the field of precessing magnetization. The latter is usually negligibly small compared to the former two. The feedback from a conducting film to the transducer is provided by the microwave electric field of precessing magnetization,[21] so that the absorbed microwave power is given by the product of the electric field at the film surface and the microwave current in the transducer [ibid].

Panels (a) and (c) in Fig. 4 show mode profiles for the first higher-order standing-wave mode for a bi-layer film which has a 90nm-thick layer of Permalloy ($4\pi M_s$=10700 Oe) and a 10nm-thick pinning layer exchange coupled to it. In the case of panel (a) the pinning layer has saturation magnetization which is smaller than that for Permalloy ($4\pi M_{s(p)}$=4000 Oe). One sees that at the boundary between the two layers $y$=90 nm the dynamic magnetization drops, and the magnetization amplitude in the pinning layer is smaller than for the bulk of the film. Furthermore, one sees that the mode profile inside the bulk is asymmetric with the maximum of amplitude located at the film surface $y$=0 facing away from the pinning layer.

Obviously, in order to excite this mode most efficiently one has to place the film with respect to the microstrip in such a way that the maximum of the mode profile faces the transducer. In this particular case the surface $y$=0 should face the transducer to meet this requirement. This will form a profile for the total microwave field shown in Fig. 4b. This profile is characterized by the maximum of the field at $y$=0 and thus by the maximum overlap integral with the mode profile in Fig.4a. The latter ensures the maximum excitation efficiency.

Furthermore, the asymmetric mode profile results in an asymmetric profile of the electric field with the maximum again at $y$=0. Thus, it is favourable to place a pick up transducer at the surface $y$=0 in order to ensure the maximum efficiency of detection for this mode. Thus, one can conclude that the maximum of efficiency of excitation of the first exchange mode for $M_{s(p)} < M_{sb}$ is obtained when the microstrip transducer is placed at the film surface facing away from the pinning layer.

Turn now to Panels c and d, describing the opposite situation $M_{s(p)} > M_{sb}$. They show the same profiles for the pining layer with saturation magnetization larger than for the bulk of the film ($4\pi M_{s(p)}$=16000 Oe). One clearly sees that in this case the mode profile (Fig 4c) is distorted in the opposite way. The maximum of the amplitude of dynamic magnetization is now in the thin layer (which actually does not pin dynamic magnetization in the bulk of the film for this particular mode. However, it does pin magnetization for the fundamental mode, as shown in Fig. 2(b) in Ref.21. ) Thus the best way to excite this mode is by placing the microstrip transducer at the surface $y$=100nm, as it will result in the total microwave magnetic field with the maximum at the same surface (see Fig.4d). Furthermore, detection of this mode is also maximized by placing the transducer at $x$=100nm, since the microwave electric field of precessing magnetization is larger at this surface (Fig.4d).). This result is in excellent agreement with experiment on Py[40-80nm]/Co[10nm] bilayer films.[24].

Thus, one can conclude that if the pinning layer has $M_s$ smaller than the bulk of the film, in order to detect its presence the microstrip transducer should be placed at the film surface which faces away from the pinning layer. If the pinning layer has $M_s$ larger than the bulk of the film, the pinning layer should face the transducer.

We also checked our findings on a considerable number of other single-layer Permalloy films of different thicknesses (30-100nm) grown in different conditions and on different setups and compared our broadband FMR results with the traditional cavity FMR measurements. All these measurements confirmed our idea: when the cavity FMR data showed a presence of a weak higher-order resonance, which is traditionally considered as evidence of magnetization pinning at film surfaces, the broadband FMR measurements for two different sample placements were able to indicate at which of the sample surfaces the magnetization pinning takes place (Fig. 5), provided we assume that the saturation magnetization for the spontaneously formed layer is smaller than for the bulk of the film. Fitting the broadband data with a theory allows one to estimate the thickness of the pinning sub-layer. If it is negligibly thin, one can describe its effect as a surface pinning, while a sub-layer of finite thickness can also pin magnetization at the interface of the bulk of the film with the sub-layer. Note that explanations alternative to a depleted layer may also exist. Examples are different anisotropy fields, large scale roughness, or weakened exchange field at a film surface. Considering them was outside the scope of this paper. The key point is that our experiment allows one to detect at which film surface this inhomogeneity is located and to extract parameters of this inhomogeneity if the inhomogeneity type is known.

To make our characterization more complete and reliable, we have also carried out numerical simulations of the structure of BLS spectra. This theoretical analysis was based on an ad hoc simplified theory, given in more detail in the appendix. Representative results for the saturating magnetic field $H$ = 1000 Oe are presented in Fig.6a (angle of incidence θ = 15°) and Fig.6b (angle of incidence θ = 30°).

It should be noted that the amplitudes of the peaks produced by the DE mode in relatively thick metal films typically exhibit a pronounced Stokes /anti-Stokes asymmetry, which is our case. Furthermore, the DE mode contributing to the Stokes peak (negative frequency shift) is confined to the top of the film. This makes the peak corresponding to negative frequency shifts more pronounced, due to a better overlapping of the three interacting waves (see the appendix). This means, in particular, that in our sample the top depleted sub-layer produces a non-negligible contribution to the BLS response. At the same time, strictly speaking, we do not know its optical parameters. Although we have neglected purely optical effects, supposing that the sample may be considered *optically uniform*, the agreement between the experiment and theory is still good.

To check the accuracy of our approximate approach, we have compared the results presented in Fig.6 with those returned by a rigorous program based on the Dissipation-Fluctuation Theorem and exact expressions for optical Green's functions.[28] The differences between the two sets of curves turned out to be insignificant. Moreover, the approximate approach provides for a better fit of the peaks produced by a hybridized SWR mode. The explanation is straightforward: this mode is very slow and therefore produces an optical response smeared in the reciprocal space. Our otherwise approximate program takes into account a finite numerical aperture while the rigorous one does not.

## Conclusion

In this paper we have demonstrated the efficiency of the broadband microstrip FMR technique for detection and characterization of a hidden magnetic inhomogeneity in a film sample. Based on the non-reciprocity of the microstrip FMR response with respect to the direction of penetration of the exciting microwave field in the sample, it is especially effective in the case when the hidden layer breaks the symmetry of the ferromagnetic structure. In the particular case of a 100 nm thick Permalloy film, an additional magnetically depleted top sub-layer, practically unidentifiable by the conventional cavity FMR setup, has been detected and characterized. These results have been confirmed by BLS spectroscopy revealing, surprisingly, the fact that the optical properties of the additional sub-layer do not differ much from those of the bulk of the film.

In a broader context, the proposed technique can be regarded as a non-destructive express method to detect the presence of magnetic inhomogeneity in conducting ferromagnetic films.

**Acknowledgment**

Support by Australian Research Council and from the National Research Foundation, Singapore under Grant No. NRF-G-CRP 2007-05 is acknowledged.

## Appendix. Simple model for BLS intensities

Here we give details of the simple model we used to calculate intensities of the BLS peaks. Since the exact theory of this effect is cumbersome and requires sophisticated numerical models,[29,30] in this work we use a simpler ad hoc approach, providing for more physical insight. It can be regarded as an extension of the formulas in Refs 31, 32. We are not interested in comparing *absolute values* of BLS intensities measured for *different* incidence angles. Our goal is to derive a tool to compare intensities for different modes observable for the same incidence angle. Therefore all calculations are made to a constant which is the same for all spin wave modes seen at the same angle of incidence.

The BLS intensities are determined by two major considerations, namely the efficiency of the magneto-optical (MO) interaction proper and that of the excitation of thermal magnons. The efficiency of the MO interaction with an individual magnon can be represented as a product of two separate factors:

$$I = I_v \cdot I_s \tag{1}$$

Here the first one

$$I_v = \left|\mathbf{e}_i \cdot (\mathbf{e}_s \times \mathbf{m})\right|^2, \tag{2}$$

called typically the vector factor, is due to the vector nature of the three-wave MO interaction and it reduces in optically isotropic media to a mixed product of the polarizations of the interacting waves: $\mathbf{e}_i$ and $\mathbf{e}_s$ are those of the incident and scattered light respectively, and $\mathbf{m}$ describes dynamic magnetization. Its physical sense is perfectly clear: the vector product, corresponding to the orientation of the polarisation induced in the film by the presence of the magnon, reflects the symmetry of the MO Faraday effect represented by the unitary anti-symmetric Levy-Civita tensor. The scalar product describes the projection of the induced polarization on the direction of polarization of the scattered wave. Strictly speaking, this scalar product implies a complex conjugate, but since in our case the scattered wave is “s” polarised $\mathbf{e}_s{}^* = \mathbf{e}_s$.

The second coefficient $I_s$, which can be called spatial factor, reflects the spatial overlapping of the afore-mentioned three interacting waves. It can be estimated, up to a constant factor, as a convolution, in $y=0$, of the induced optical polarization $\exp(-ik_y y)\cdot m(y)$ and the Green’s function $\exp(-ik_y\left|y-y'\right|)$, the latter being an optical response, in the backscattering geometry, to a delta layer of polarization $\delta(y-y')$:

$$I_s = \left|\int_L \exp(-2ik_y y')m(y')dy'\right|^2, \tag{3}$$

where $k_z = (q+ip)\dfrac{2\pi}{\lambda}$ is the complex out-of plane wave number ($p$<0) and λ = 530 nm is the laser wavelength.

The simplified expression (3) is justified because in our case the thickness $L$ of the samples studied in this work is considerably larger than the skin depth at the laser wavelength. That is why one can neglect the interference created by the reflection from the interface $y= L$. At the same time, it should be noted that the approximate expression (1), representing the efficiency as a product of two independent factors, holds only in the case where the spin mode polarization does not depend on $y$. The latter is correct only for the case of non-hybridized modes. However, if a pronounced hybridization takes place the vector factor $I_v$ should kept under the integral. (Since our experimental data suggest mode hybridization, in our numerical calculation we actually used the latter approach.)

To obtain the final formula we also take into account the fact that the lens collects scattered light in the range from $\theta - \Delta\theta/2$ to $\theta + \Delta\theta/2$. This results in the aperture factor (see Eq.(36) in Ref.32

$$I_a = 2\,\mathrm{atan}(V_g \delta k / \omega'')/(V_g \omega''), \quad (4)$$

where $V_g$ is spin-wave group velocity and ω'' is the imaginary part of spin wave group velocity responsible for spin wave damping. (We calculate this factor for the maximum of the BLS peak.) The uncertainty in the transferred wave number $\delta k$ relates to the collection angle via $\delta k = 4\pi \cos(\theta)\Delta\theta / \lambda$.

We also add a factor $I_{fd}$ which arises from the fluctuation dissipation theorem and which relates amplitudes of thermally excited waves to the Rayleigh-Jeans distribution:

$$I_{fd} = (\omega'')^2 / \omega', \quad (5)$$

where ω' is the real part of the complex spin wave eigenfrequency. Finally, the scattered intensity is proportional to the product of the factors (2)-(5):

$$I \sim I_v \cdot I_s \cdot I_{fd}. \quad (6)$$

This expression was used in this work to calculate BLS intensities shown in Fig.6. For each of the layers its own value for Gilbert damping parameter was used (0.008 for Permalloy and 0.16 for the depleted layer.) By solving the eigenvalue-eigenvector problem for the dipole-exchange operator[34] for the bi-layer complex eigenfrequencies $\omega' + i\omega''$, group velocities, and modal profiles for all spin wave modes were obtained. These data were inserted into (6) to yield the intensities.

## Figure captions

Fig. 1. Broadband FMR responses for a 100-nm Permalloy film showing evidence of the presence of a magnetically depleted layer. Driving frequency is 10 GHz. Solid line: film faces the microstrip transducer, dashed line: substrate faces the transducer.

Fig. 2. Exemplary theoretical fits of experimental data with theory from Ref.21 for different frequencies. (a): 8 GHz, substrate faces the transducer, (b) 12 GHz, substrate faces the transducer, (c) 15 GHz, film faces the transducer. Red solid line lines: experiment, blue dashed lines: theory.

Parameters of calculation.

Total film thickness: 100 nm.

Bulk of the film (90nm-thick): Saturation magnetization: $4\pi M_s$=10700G, exchange constant: $1.2\cdot10^{-6}$ erg/cm, conductivity: $4.5\cdot10^{6}$ Siemens/m. Spontaneously formed layer: Thickness: 10 nm, saturation magnetization: $4\pi M_s$=4000G, exchange constant: $0.51\cdot10^{-6}$ erg/cm, conductivity is the same as for the bulk. Inter-layer exchange constant: 60 erg/cm$^2$. The thin layer is formed at the free surface of the film.

Fig.3. Brillouin light scattering data. Different frequencies for the Stokes (black triangles down) and the anti-Stokes (red triangles up) are clearly seen. Red solid line: calculated spin wave dispersion for the "Stokes" direction of spin wave propagation. Dashed black line: the same for the opposite ("anti-Stokes" propagation direction). Blue dotted line: dispersion for a single-layer film with the same total thickness.

Fig. 4. Calculated mode profiles for the first higher-order mode for two different values for saturation magnetization of the pinning layer $4\pi M_{s(p)}$. (a) and (b): $4\pi M_{s(p)}$ =4000 KG; (c) and (d): $4\pi M_{s(p)}$ =16000 KG. The other calculation parameters are the same as for Fig. 2.
(a) and (c): red solid lines: amplitude of the in-plane component of dynamic magnetization. Blue dashed lines: its phase. (b) and (d): red solid lines: microwave electric field; blue dotted lines: total microwave magnetic field. The corresponding external microwave field is applied from the surface which ensures maximum efficiency of excitation for this mode. All unspecified measurement units are arbitrary.

Fig.5. Comparison of the broadband FMR data with the cavity FMR data for a single-layer 100-nm Permalloy film grown on a sapphire substrate and having 10nm-thick gold capping and seed layers. Blue dashed line: film facing the broadband transducer. Red dotted line: substrate facing the transducer. Black solid line: cavity FMR data. In order to make a valid comparison, in all three cases a microwave diode, sine wave modulation of the applied field and lock-in technique was used to detect absorption. The graph shows anti-derivative of the raw data. The data indicate a presence of magnetization pinning at the film interface with the gold capping layer. Frequency: 9.527 GHz.

Fig. 6. Measured BLS intensities and their fits with the theory in the Appendix. Parameters of calculation are the same as for Fig. 2. The laser light is incident from the side of the magnetically depleted layer. a) light incidence angle θ=15° b) θ=30° , applied field $H$= 1000 Oe.

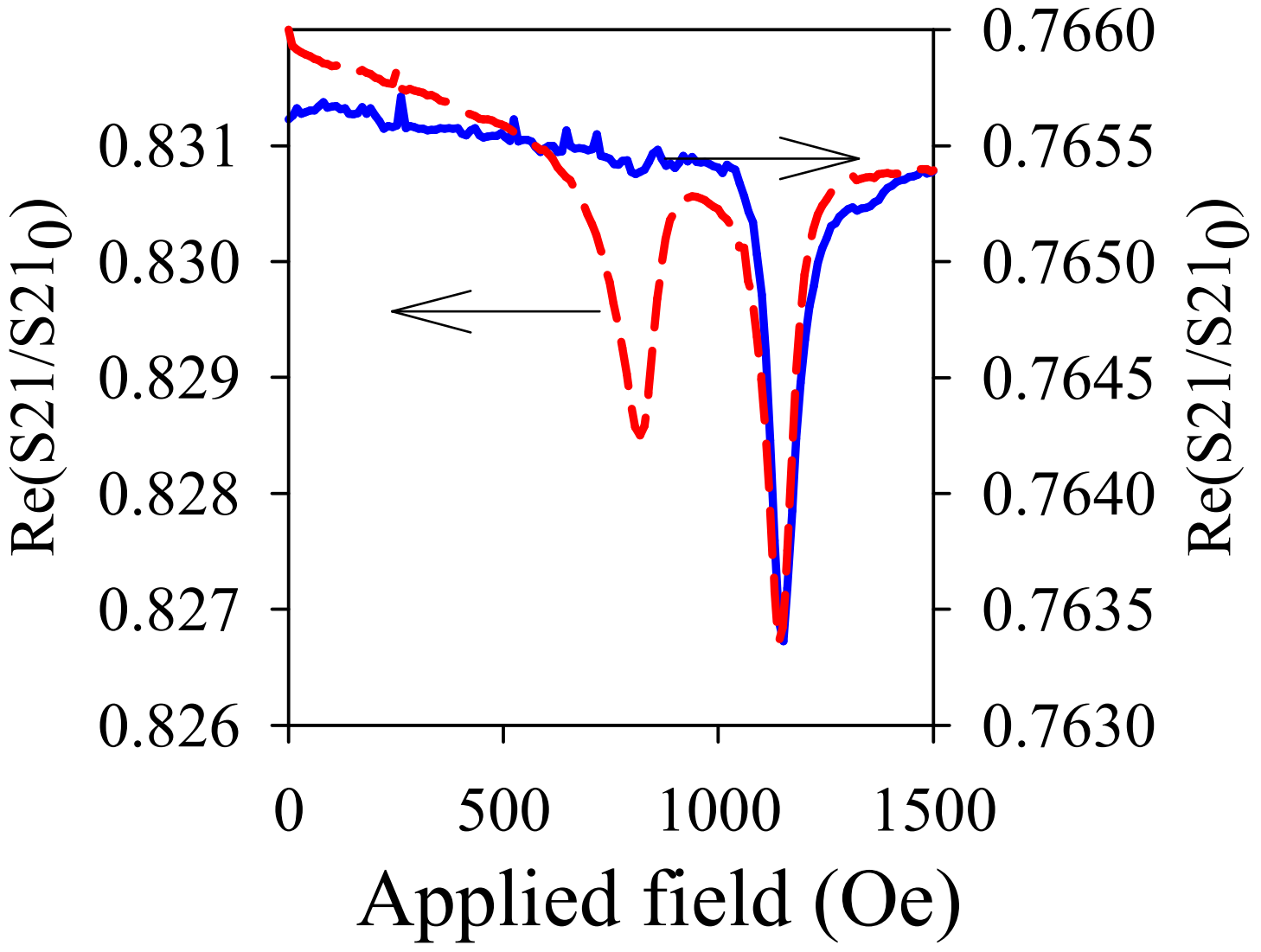

Fig. 1

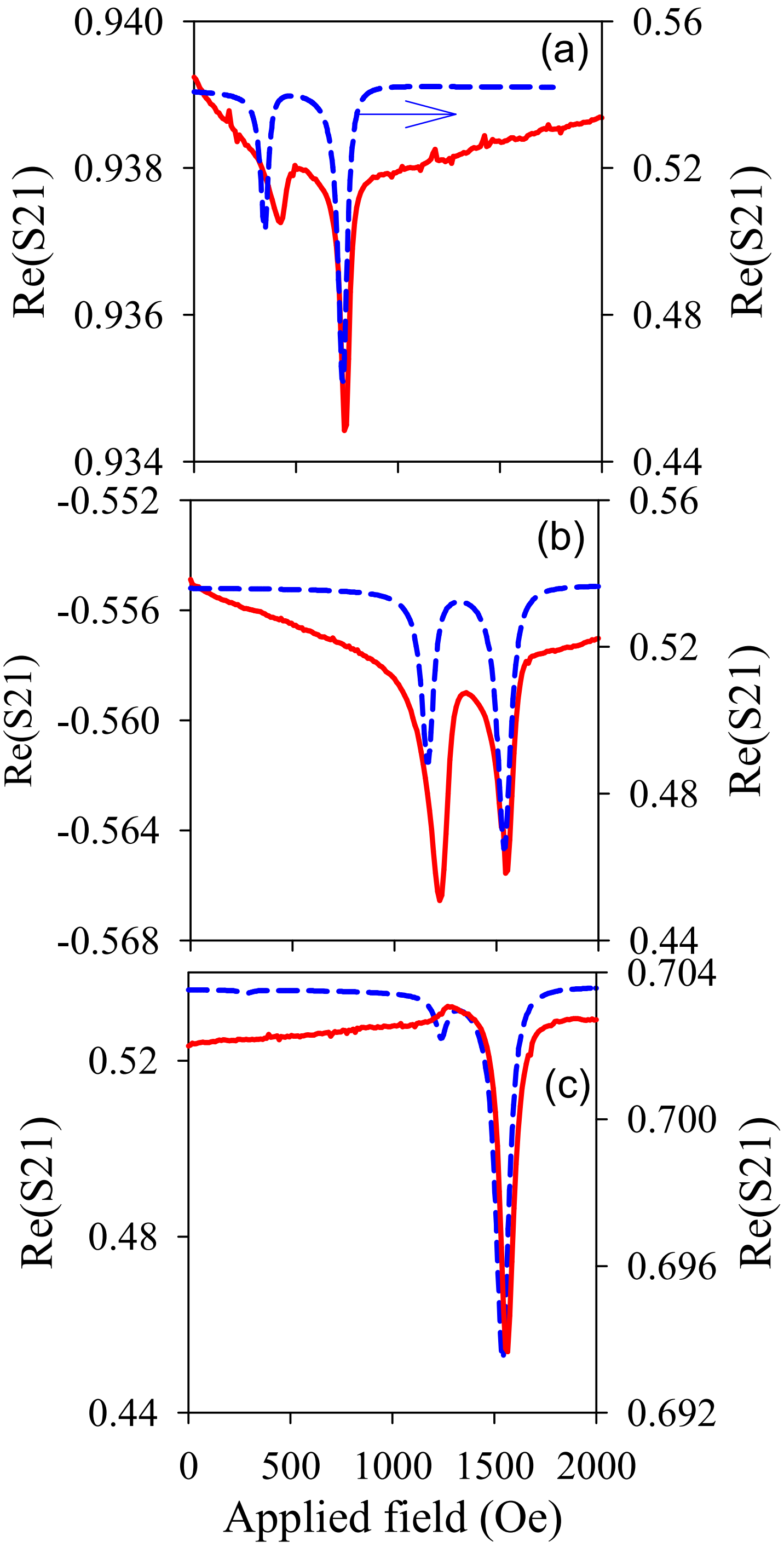

Fig. 2

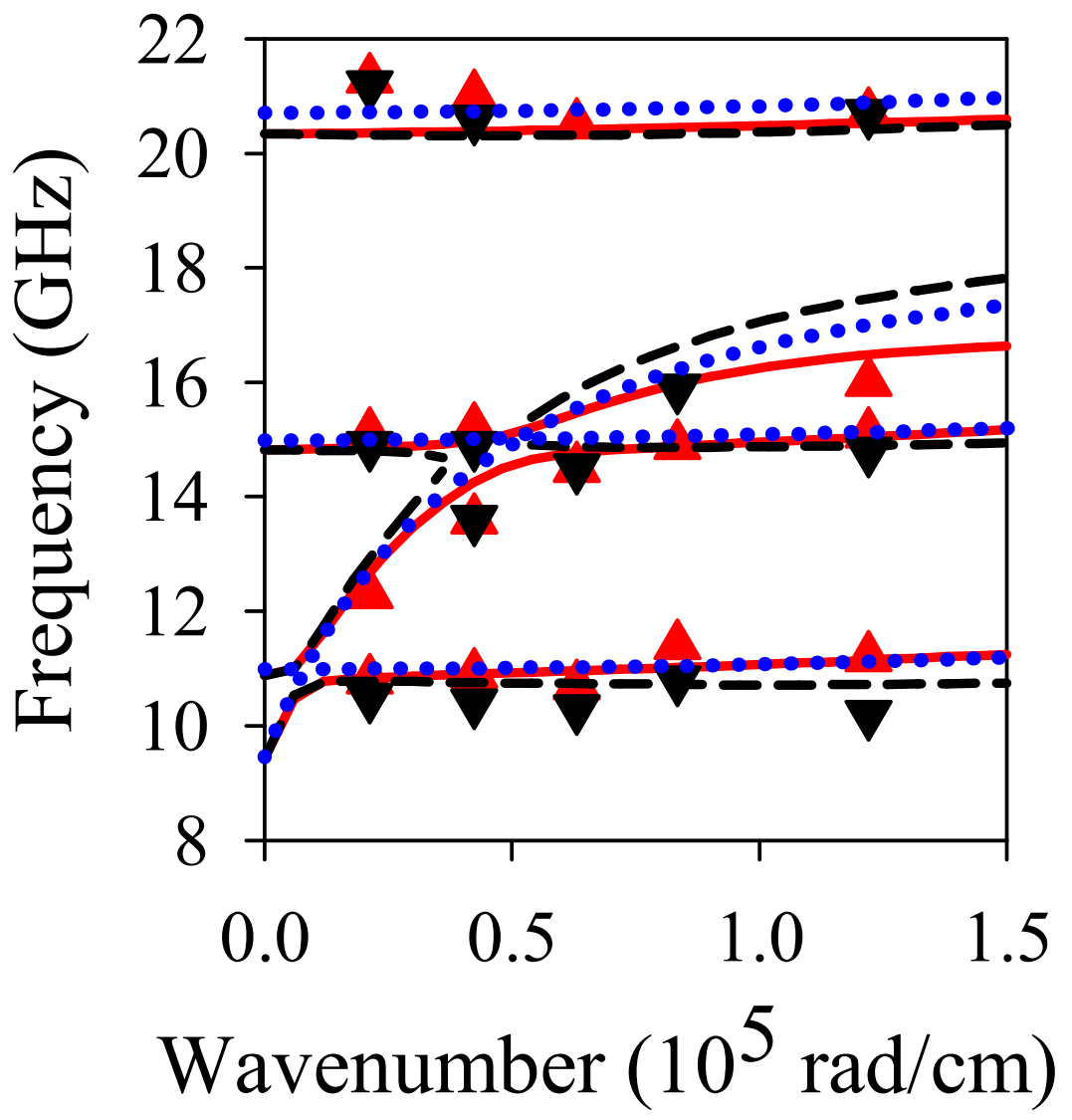

Fig. 3

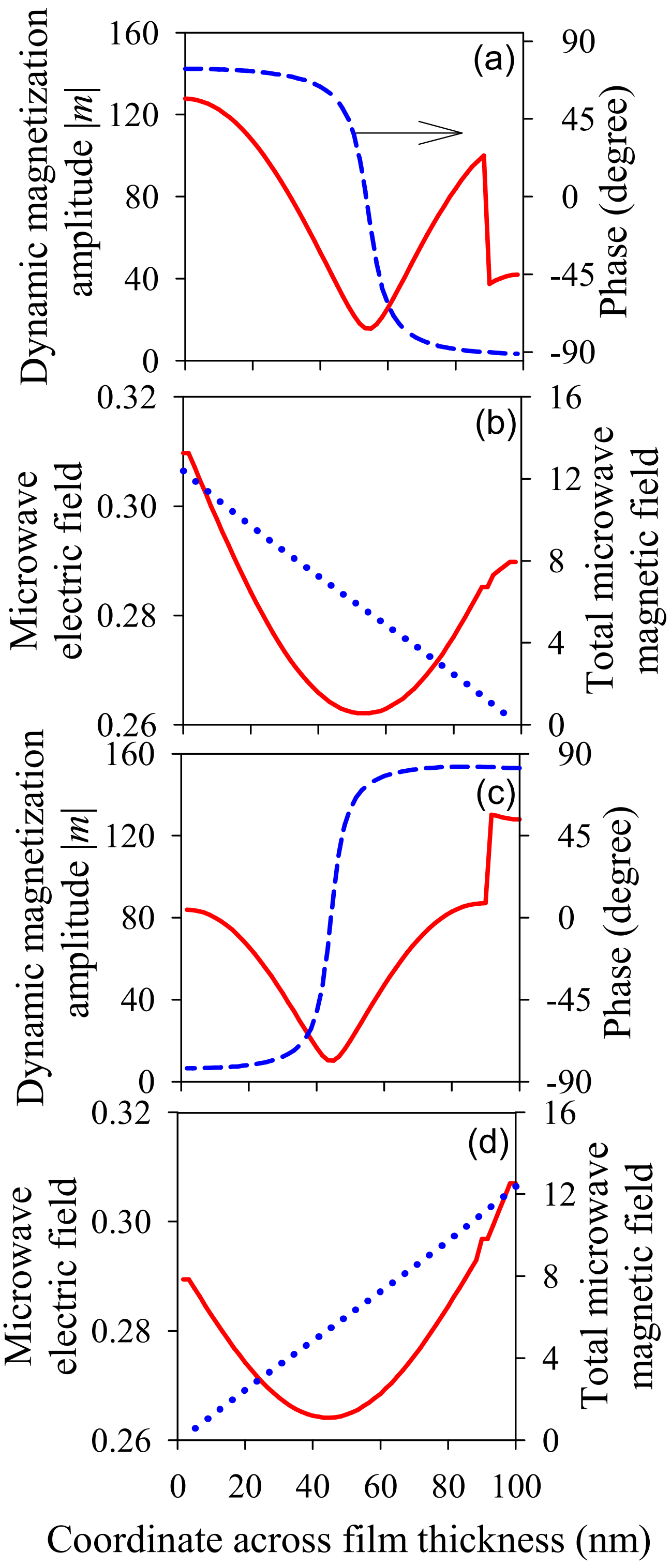

Fig. 4

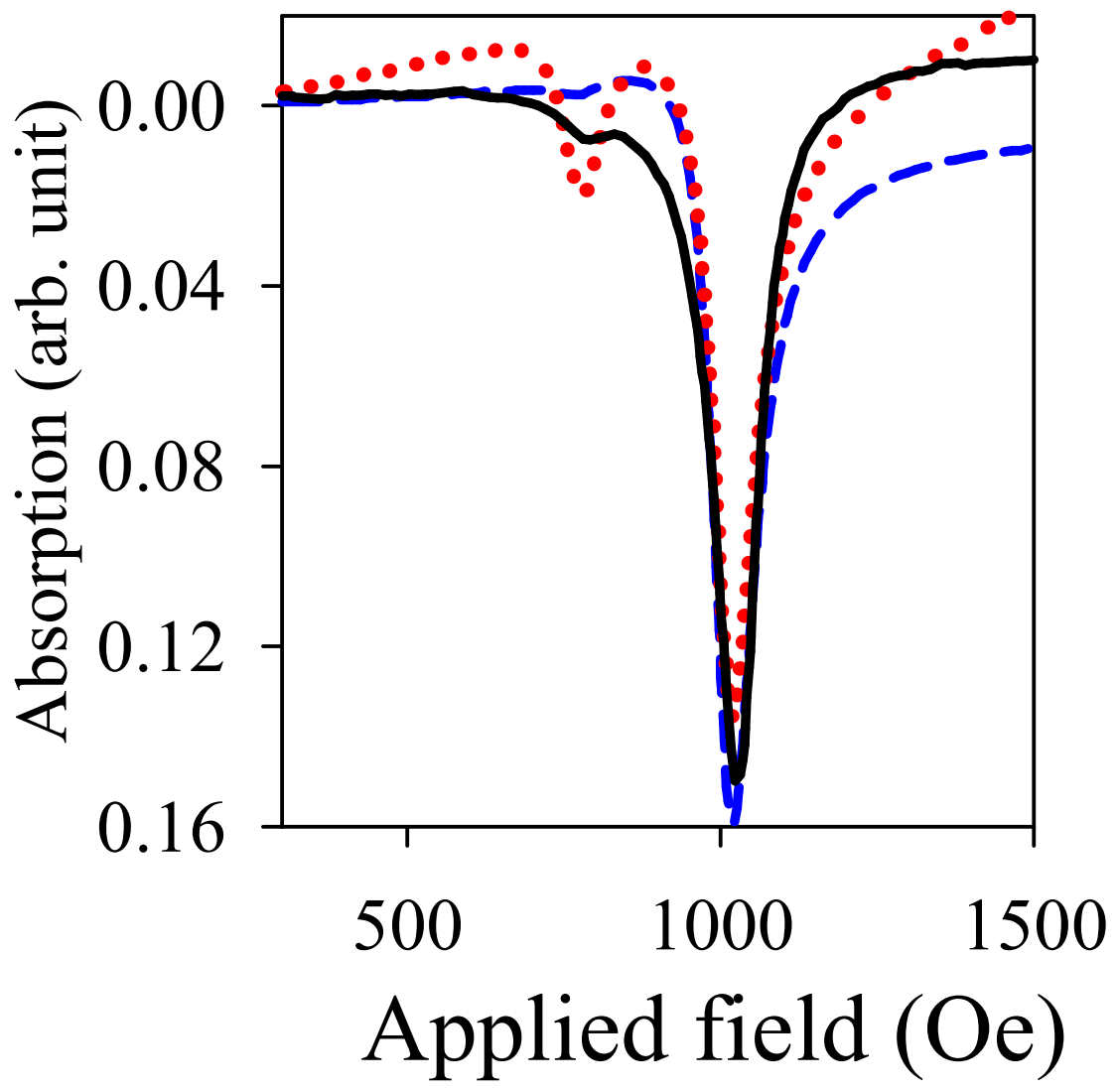

Fig. 5

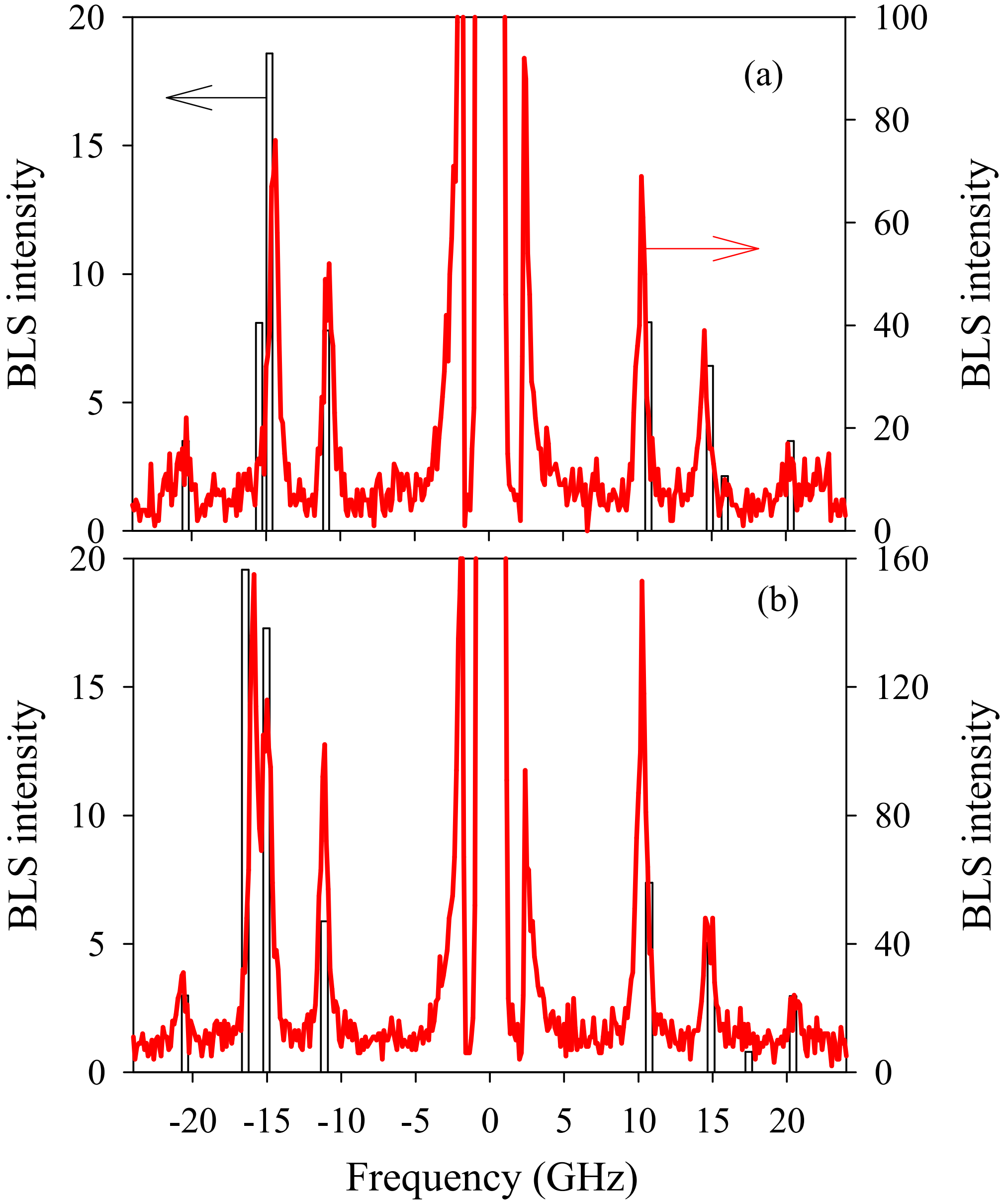

Fig. 6